\documentclass[]{ceurart}

\usepackage{ragged2e}
\begin{document}
\copyrightyear{2022}
\copyrightclause{Copyright for this paper by its authors.
  Use permitted under Creative Commons License Attribution 4.0
  International (CC BY 4.0).}

\conference{Forum for Information Retrieval Evaluation, December 9-13, 2022, India}

\title{Identification of the Relevance of Comments in Codes Using Bag of Words and Transformer Based Models}

\author[1]{Sruthi S}[%
email=sruthisudheer1214@gmail.com,
]
\cormark[1]
\cortext[1]{Corresponding author.}
\address[1]{Department of Data Science and Engineering, Indian Institute of Science Education and Research, Bhopal}

\author[1]{Tanmay Basu}[
email=welcometanmay@gmail.com,
url= https://sites.google.com/view/tanmaybasu/,
]


\begin{abstract}
The Forum for Information Retrieval (FIRE) started a shared task this year for classification of comments
of different code segments. This is binary text classification task where the objective is to identify whether comments given for certain code segments are relevant or not. The BioNLP-IISERB group at the Indian
Institute of Science Education and Research Bhopal (IISERB) participated in this task and submitted
five runs for five different models. The paper presents the overview of the models and other significant
findings on the training corpus. The methods involve different feature engineering schemes and text
classification techniques. The performance of the classical bag of words model and transformer-based
models were explored to identify significant features from the given training corpus. We have explored
different classifiers viz., random forest, support vector machine and logistic regression using the bag of
words model. Furthermore, the pre-trained transformer based models like BERT, RoBERT and ALBERT
were also used by fine-tuning them on the given training corpus. The performance of different such
models over the training corpus were reported and the best five models were implemented on the given
test corpus. The empirical results show that the bag of words model outperforms the transformer based
models, however, the performance of our runs are not reasonably well in both training and test corpus. This paper also addresses the limitations of the models and scope for further improvement.
\end{abstract}


\maketitle
\section{Introduction}\label{introduction}
A crucial part in the software development and management industry is program comprehension. It is a usual practice to add new functionalities to the existing programs to meet the dynamic requirements. Therefore understanding the relevant parts and functions of the code is extremely important to avoid redundancy and inefficiency \cite{10.1145/1085313.1085331}. Developers often act themselves as end users to identify any malfunctions \cite{10.5555/2337223.2337254} and to mine the process. Even though qualified personnel or peer code reviewer can handle this cumbersome process, the resources needed are quite high. Over the years, researchers formalized structured process for code inspection, the need for inspection and debugging and its benefits \cite{5388086}. Industries like Microsoft, Google and some open source platforms use another version of the peer code review supported with tool-based approaches known as modern code review or contemporary code review \cite{bird2013expectations}. The most useful block in a program review are comments. They are direct and descriptive rather than the source code. This helps reviewers to analyse, add feedback and suggestions  which, in a sense, influence the effectiveness of such practices. Nevertheless, some comments can be irrelevant or redundant which make this task even more complicated. So assessing the quality of comments is also necessary as they act as a guide to the reviewer \cite{Majumdar2020}.  \\

The Forum for Information Retrieval (FIRE) 2022 started a shared-task this year, named as Information Retrieval in Software Engineering (IRSE)\cite{majumdar2022overview}, to evaluate the relevance of comments with respect to its surrounding codes. The objective of this shared task is to build a reusable benchmark for evaluating the models to classify the relevance of the comments of given source codes \cite{majumdar2022overview}. This is a binary text classification problem to categorize the source code comments as useful or not useful, where the given codes were written in C programming language. We, the BioNLP research group at Indian Institute of Science Education and Research Bhopal (IISERB) participated in this shared task and explored the performance of various feature engineering and classification techniques to categorize the source code comments. \\ 

The proposed framework generates features from the given training corpus by using the classical bag of words (BOW) model \cite{manning_raghavan_schütze_2008} and transformer architecture based deep learning models \cite{bert,roberta,Lan2020ALBERTAL}. The term frequency (TF) and inverse document frequency (IDF) i.e., TF-IDF \cite{manning_raghavan_schütze_2008} and Entropy based method \cite{sabbah17} were used as the term weighting schemes. Eventually, the performance of the classifiers like support vector machine, logistic regression and random forest that uses BOW features on the training corpus have been reported . Furthermore, three different attention layer based deep learning models, viz. BERT \cite{bert}, ALBERT\cite{Lan2020ALBERTAL}, and RoBERTa\cite{roberta} were implemented to generate semantic features from the given training corpus and then those features were used for comments classification. The top five frameworks that have best performance on the training corpus were chosen based on the F1 score and accuracy. Consequently, these models were implemented on the test data and submitted. \\  

Section \ref{literature} of  the paper presents the related works in code comment classification. Section \ref{expdes} describes the proposed frameworks. The experimental results are recorded and analyzed in section \ref{result}. Ultimately, the work is concluded in section \ref{conclusion}. 

\section{Related Works}\label{literature}
This section briefly describes the related works in developing an architecture that can be used for comment classification. As comments are written in natural language while codes are programming language, detecting inconsistencies of codes with comments are often difficult. Tan et al.  \cite{10.1145/1323293.1294276} developed a framework known as iComment which combines Natural Language Processing (NLP), Machine Learning, Statistics and Program Analysis techniques to overcome this issue. They experimented on four large code bases, namely, Linux, Mozilla, Wine and Apache. The framework has an accuracy of  90.8-100\% which also detects bad comments. \\

Similarly, the base research work of this challenge, named as Commentprobe \cite{https://doi.org/10.1002/smr.2463}, was implemented on C codebases. First, a developer survey to study the commenting behaviour among the programmers was conducted and then the ground truth for the comment classification task was generated by manual annotation. They developed pretrained embeddings known as SWVec using the data from the posts in Stack Overflow and literature works. These features are trained using neural network architecture like LSTM and ANN to classify the comments as useful or not, or partially useful and could achieve an F1 score of  86.34\%. \\

Apart from the comment classification, many other attempts have been done to understand and transform the code of different programming languages. Some of the works on different types of embeddings are explored in the paper \textit{A Literature Study of Embeddings on Source Code} \cite{https://doi.org/10.48550/arxiv.1904.03061}.

\section{Experimental Design} \label{expdes}
The given training and test corpora contain comments, corresponding C code snippets and the class labels, which denotes whether the given comments are useful or not useful. Both the training and test data have these three information and were  released in csv format. We extracted the codes and comments to make two different corpora as one containing only comments and the other containing both the code and comments. 

The logistic regression (LR) \cite{lr}, random forest (RF) \cite{rf} and support vector machine (SVM)\cite{svm1,basu16} classifiers were implemented using both TF-IDF based term weighting scheme \cite{manning_raghavan_schütze_2008} and Entropy based term weighting scheme \cite{basu21} following the bag of words model. In Entropy based term weighting scheme, the weight\footnote{\url{https://radimrehurek.com/gensim/models/logentropy\_model.html}} of a term in a document is determined by the entropy of term frequency of the term in that document \cite{sabbah17,basu21}. We implemented the $\chi^{2}$-statistic and mutual information \cite{trl16} based term selection techniques to identify a predefined number of top terms from the bag of words. We had done the experiments by using different numbers as threshold for the $\chi^{2}$-statistic and mutual information based term selection method and then reported the best result for each model. These models were trained only using the comments of the given training corpus and were implemented in Scikit-learn\footnote{\url{http://scikit-learn.org/stable/supervised\_learning.html}}, a ML tool in Python. The parameters of the classifiers were tuned using 10-fold cross validation scheme on the training corpus. We did not use the codes to train these models as the BOW model cannot identify relevant characteristics of the code snippets. 

Moreover pre-trained transformer based models viz. BERT\footnote{\url{https://huggingface.co/bert-base-uncased}} \cite{bert}, RoBERT\footnote{\url{https://huggingface.co/roberta-base}} \cite{roberta} and ALBERT\footnote{\url{https://huggingface.co/albert-base-v1}} \cite{Lan2020ALBERTAL} were used and fine-tuned on the given training corpus using both the code and comments following 10-fold cross validation scheme. For ALBERT and ROBERTa models, the length of the tokenized text is fixed as 432 and were trained over 18 and 38 epochs respectively. Subsequently, the best setting of each of these models were tested on the test corpus. The top five models which perform better than the other on the test corpus were submitted to the organizers for final evaluation. The code and data set that are used to implement the proposed framework are available on Github\footnote{\href{https://github.com/SruthiSudheer/Comment-classification-of-C-code}{https://github.com/SruthiSudheer/Comment-classification-of-C-code}}. 

\begin{table}
\centering
\caption{Performance of Different Frameworks on the Training Corpus} \label{results_training}
\setlength{\tabcolsep}{4pt}
\begin{tabular}{|c|l|c|c|c|c|}
\hline
Feature Types & Classifier  & Accuracy & Precision  & Recall & F1 Score\\
\hline

  & Logistic Regression    &0.66 &0.71	&0.62&	0.67  \\
\textbf{Entropy Based Features}    & Random Forest       &0.68   &0.73	&0.65	&0.69 \\
(using only comments)                        & Support Vector Machine &0.67 &0.72	&0.63   &0.67  \\
\hline
                       
   & Logistic Regression    &0.69 &0.71	&0.73&	\textbf{0.72}	\\
\textbf{TF-IDF Based Features}      & Random Forest       &0.67   &0.72	&0.65   &0.68  \\
(using only comments)                        & Support Vector Machine &0.69 &0.71	&0.70	&\textbf{0.71}  \\
\hline
                        
 & BERT &	0.54 &0.56	&0.66	&	0.61\\
\textbf{Transformer Based Features}          & RoBERTa              &0.52  & 0.54	&0.84	&\textbf{0.66}	 \\
(using both code and comments)                  & ALBERT &0.53  &0.54	&0.87	&\textbf{0.67}	        \\
		                                             
\hline
\end{tabular} 
\end{table}

\section{Results and Analysis}\label{result}
Table \ref{results_training} shows the performance of different models on the training corpus. It may be noted from table \ref{results_training} that the logistic regression (LR) and SVM classifiers using the classical TF-IDF based term weighting scheme of the bag of words model respectively achieves the best and the second best F1 scores among all the other models. The ALBERT model outperformed BERT and ROBERTa models in terms of F1 score, however, they could not beat the LR and SVM classifiers. Table\ref{results_test} shows the performance of the best five models on the test corpus. Note that we had run the best setting of all the models individually on the test corpus, but just reported the results of the best five models that we submitted as our final runs. The best settings of individual frameworks are also reported in Table\ref{results_test}. It can be observed from Table\ref{results_test} that almost all the models achieve poor results on the test corpus in comparison to their performance on the training corpus. 
\begin{table}
\centering
\caption{Performance of the selected Frameworks on the Test Corpus} \label{results_test}
\setlength{\tabcolsep}{3pt}
\begin{tabular}{|c|l|c|c|c|c|c|}
\hline
Submitted & Framework &Significant  & Accuracy & Precision  & Recall & F1 Score\\
Results  &   &Parameters   &   &   &  & \\
\hline
 Run 1 &TF-IDF + RF   &sm$^{7}=$information gain, \#trees = 50,  &0.47 &0.34	&0.97&	0.51	\\
  &   & \#terms=3000, $\chi^{2}$-statistic  &   &   &  & \\
 \hline
 Run 2  & Entopy+SVM &linear kernel, $C^{8}=$1, $\gamma=$scale,  &0.53 &0.37	&0.94	&0.53	 \\ 
  &   & \#terms=3000, $\chi^{2}$-statistic  &   &   &  & \\
 \hline
Run 3 &Entropy+RF    &sm$^{7}=$information gain, \#trees = 50  &0.41 &0.32	&0.98	&0.49	 \\
 &   & \#terms=3000, $\chi^{2}$-statistic  &   &   &  & \\
\hline
 Run 4 & ALBERT &epochs=18,ws$^{9}=$500,len$^{10}=$432   &0.57 &0.33	&0.48&	0.39	      \\    
 &&batch size=4,weight decay=0.01   &   &   &   &   \\
\hline
Run 5 &RoBERTa    &  epochs=38,ws$^{9}=$500, len$^{10}=$432   &0.58      &0.32	&0.42&0.36	\\
 &&batch size=4,  weight decay=0.01     &   &   &   &   \\
\hline
\end{tabular} 
\RaggedRight \quad \footnote *Splitting measure. \footnote *Cost parameter. \footnote *Number of warmup steps for learning rate scheduler. \footnote*Length of tokenized text
\end{table}

The entropy based SVM classifier outperforms the other models on test corpus, however, it could not beat the performance of many classifiers using the training corpus. On the other hand, the LR and SVM classifiers using TF-IDF based bag of words features, which performed very well on the training corpus, did not produce a similar performance on the test corpus. We could not find the reasons behind such poor performance of many such models due to time constraint, but in future we plan to investigate the same. All the transformer based models perform poorly on the test corpus. The major reason behind this may be the semantics that were necessary for the comments of the test corpus were not captured during training stage by the transformer based models as the size of the training corpus was insufficient for such models. Moreover, the the pre-trained models that we used were developed using the Books corpus\footnote{\href{https://huggingface.co/datasets/bookcorpus}{https://huggingface.co/datasets/bookcorpus}} and Wikipedia\footnote{\href{https://huggingface.co/datasets/wikipedia}{https://huggingface.co/datasets/wikipedia}} and hence they could not capture the semantics of the codes from the comments. 

\section{Conclusion}\label{conclusion}
The task offered by IRSE Track on FIRE 2022 highlights various challenges for identifying useful comments and thus removing redundancy and non dependency of comments with the source code. In this perspective, we have implemented different frameworks using various types of text features from the given training corpus to identify the relevance of code comments. From the perspective of empirical analysis none of the models achieve reasonable performance on the test corpus. In future, we need to investigate the reasons to develop novel models to improve the performance. 
However, we feel that the given training corpus of codes and comments are very small in size and hence cannot capture all the behavioural aspects of code and comments. We barely used software development concepts which indeed, could have developed a new embedding that can significantly identify relevant features useful in the software domain. The proposed text classification based approaches could not capture all the necessary semantics for software development and maintenance, which needs to be addressed in future.

\bibstyle{unsrt}
\bibliography{irse2022}

\end{document}